\documentclass[fleqn,twoside]{article}
\usepackage{espcrc2}
\def\lesssim{\lower.7ex\hbox{${\buildrel < \over \sim}$}}
\def\gtrsim{\lower.7ex\hbox{${\buildrel > \over \sim}$}}
\def\mp{\lower.7ex\hbox{${\buildrel - \over +}$}}

\usepackage{graphicx}

\title{Neutrino production in UHECR proton interactions in the infrared background}
\author{Todor Stanev\\
Bartol Research Institute, University of Delaware\\
Newark, DE 19716 USA}
\begin{document}
\begin{abstract}
 We discuss the contribution of proton photoproduction interactions
 on the  isotropic infrared/optical background to the cosmic neutrino
 fluxes. This contribution has a strong dependence on the proton
 injection energy spectrum, and is essential at high redshifts. 
 It is thus closely correlated with the cosmological evolution of 
 the ultra high energy proton sources and of the inrared background
 itself. These interactions may also contribute to the source
 fluxes of neutrinos if the proton sources are located in regious 
 of high infrared emission and magnetic fields. 

\vspace{1pc}
\end{abstract}
\maketitle


 The assumption that the Ultra High Energy Cosmic Rays (UHECR) are
 nuclei (presumably protons) accelerated in luminous extragalactic
 sources provides a natural connection between these particles and
 ultra high energy neutrinos. This was first realized by
 Berezinsky\&Zatsepin~\cite{BerZat69} soon after the introduction of
 the GZK effect~\cite{GZK}. The first  realistic calculation of the
 generated neutrino flux was made by  Stecker~\cite{Stecker73}.
 The problem has been revisited many times after the paper of
 Hill\&Schramm~\cite{HS85} who used the non-detection of such neutrinos
 to limit the cosmological evolution of the sources of UHECR. 
 
 These so called cosmological neutrinos are produced in photoproduction
 interactions of the UHECR with the ambient photon fields, mostly with
 the microwave background radiation (MBR). The GZK effect is the limit
 on the highest energy a cosmic ray proton can retain in propagation
 through the MBR.  It sets a cutoff in the cosmic ray energy spectrum
 in case the UHECR sources are isotropically and homogeneously
 distributed in the Universe. The physics of these
 photoproduction interactions is very well known. Although the energy
 of the interacting protons is very high, the center of mass energy
 is low, mostly at the photoproduction threshold. The interaction
 cross section is studied at accelerators and is very well known.
 Most of the interactions happen at the $\Delta^+$ resonance where
 the cross section reaches 500$\mu$b. The mean free path reaches a
 minimum of 3.4 megaparsecs (Mpc) at proton energy of 6$\times$10$^{20}$ eV.
 The average energy loss of 10$^{20}$ protons is about 20\% per
 interaction and slowly increases with the proton (and center of mass)
 energy.

 The fluxes of cosmological neutrinos are, however, very uncertain
 because of the lack of certainty in the astrophysical input.
 The main parameters that define the magnitude and the spectral shape
 of the cosmological neutrino fluxes are: the total UHECR source
 luminosity $L_{CR}$, the shape of the UHECR injection spectrum
 $\alpha_{CR}$ in the case of power law spectrum, the maximum UHECR
 energy at acceleration $E_{max}$ and the cosmological evolution of
 the UHECR sources. These are the same parameters that
 Waxman\&Bahcall~\cite{WB1} used to set a limit on the neutrino
 fluxes generated in optically thin sources of UHECR.

 The microwave background is not the only universal  photon
 field that has to be taken in consideration. Especially interesting
 is the isotropic infrared and optical background (IRB). The number
 density of IRB is smaller than that of MBR by more that two orders
 of magnitude. On the other hand,  protons of lower energy can
 interact on the IRB, and the smaller number density has to be
 weighted with the larger flux of interacting protons. The present
 Universe is optically thin to 10$^{19}$ eV and lower energy
 protons, but even at small redshift the proton interaction rate
 quickly increases. This is different from the interactions on MBR,
 where the interacting protons quickly lose their energy even at
 $z$=0. The cosmological evolution of UHECR injection is thus of major
 importance for the contribution of such interactions to the flux of
 cosmological neutrinos.
  
\begin{figure}[thb]
\centerline{\includegraphics[width=3.truein]{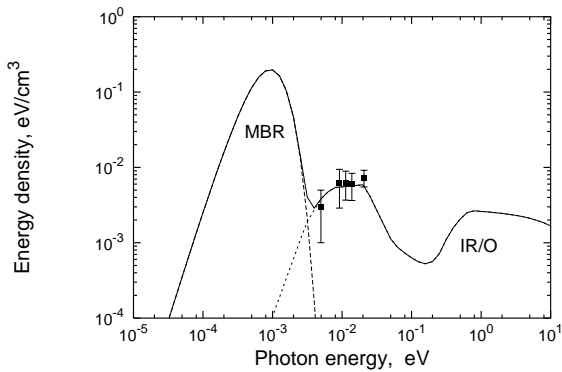}}
%
\caption{ The energy density in MBR and IRB according to the model
 of Ref.~\protect\cite{Franceschini01}. The data points are from
 analyses of the
 DIRBE measurements~\protect\cite{hauser98,lagache99}.
\label{irbmbr}
}
\end{figure}

 We use the IRB model of Franceschini et al~\cite{Franceschini01} 
 shown in Fig.~\ref{irbmbr} together with the MBR in terms of
 energy density. The model consists of two components: `star', 
 near infrared, which covers the higher photon energies, and
 `dust', far infrared that continues down to MBR. The total
 IRB number density is significantly smaller than that of MBR.
 The model yields 1.6 photons/cm$^3$, a factor of 250 less than
 the MBR. The IRB is measured directly after subtraction of
 point sources and is also estimated from the absorption of
 TeV photons coming from extragalactic sources~\cite{Steckers}.
 These estimates affect mostly the near infrared part of the
 spectrum. Photons of wavelength above 40 $\mu$m affect only
 the $\gamma$-ray fluxes above 10 TeV~\cite{StaFra} where the
 statistics is usually low and the flux decrease could also be
 due to absorption in the $\gamma$-ray sources.
 
\begin{figure}[thb]
\centerline{\includegraphics[width=3.truein]{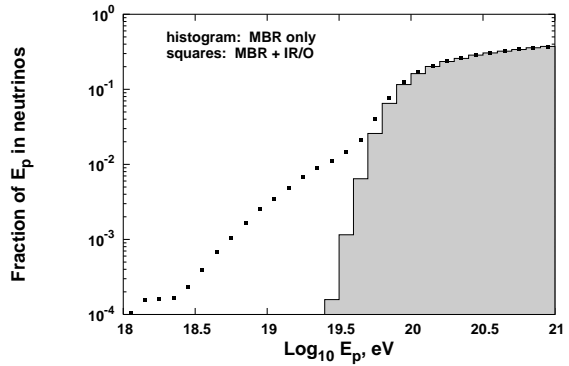}}
%
\caption{ Fraction of the proton energy that is converted
 to neutrinos on propagation on 200 Mpc. The histogram shows
 interaction on MBR only, while the points represent 
 interactions of the photon spectrum shown in Fig.~\protect\ref{irbmbr}.
\label{nuprod}
}
\end{figure}

  In addition to the lower total photon density the IRB covers
 much wider wavelength range than the microwave background, and
 its photon density per unit energy is even smaller.
 The interactions of UHECR on IRB photons are indeed very rare
 in the present universe.
 Fig.~\ref{nuprod} shows the fraction of the proton energy
 that is converted to neutrinos as a function of the proton
 energy in propagation on a distance of 200 Mpc.

 In the derivation of the neutrino limit Waxman\&Bahcall use cosmic
 ray source luminosity $L_{CR} =
 4.5 \pm 1.5 \times {\rm 10}^{44}$ erg/Mpc$^3$/yr between
 10$^{19}$ and 10$^{21}$ eV for power law cosmic ray energy spectrum
 with $\alpha$ = 2. The assumption is that no cosmic rays are
 accelerated above 10$^{21}$ eV. The cosmological evolution of the
 source luminosity is assumed to be $(1+z)^3$ to $z$ = 1.9 then
 flat to $z$=2.7  with an exponential decay at larger redshifts.
 We will first use the parameters of this limit to find the
 contribution of the proton interactions on IRB.

\begin{figure}[thb]
\centerline{\includegraphics[width=3.truein]{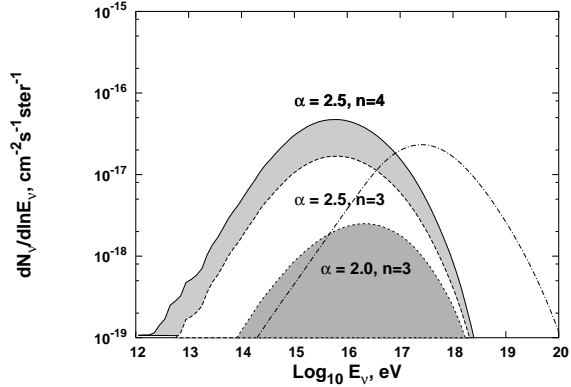}}
%
\caption{ Fluxes of cosmological neutrinos (\protect$\nu_\mu + \bar{\nu}_\mu$)
 generated only by
 interactions on IRB. All three calculations use the UHECR
 luminosity derived by Waxman~\protect\cite{Wax95}.
 The power law spectral indices and the cosmological evolution
 of UHECR sources \protect$n$ are given by each curve.
 The dash-dotted line shows the `standard' \protect$n$=3 cosmological
 neutrino flux from  interactions in the MBR.
\label{ir_only}
}
\end{figure}

 The resulting $\nu_\mu + \bar{\nu}_\mu$ spectrum for a cosmological
 model with  $\Omega_\Lambda$ = 0.7, $\Omega_M$ = 0.3 and $H_0$ =
 75 km/s/Mpc is shown with a dotted line in Fig.~\ref{ir_only}.
 The flux peaks at 10$^{16.3}$ eV at 2.5$\times$10$^{-18}$ 
 cm$^{-2}$s$^{-1}$ster$^{-1}$. The peak is at energy lower than
 the peak of the MBR interactions (shown with a dash-dot line)
 by a factor of 20, and its magnitude is also  lower by a factor
 of 10. Next we show in the same figure with a dashed
 line the contribution of IRB for a scenario in which the
 injection spectral index is changed to $\alpha$ = 2.5 and all
 other parameters are the same. There is a noticeable shift of
 the peak position to still lower energy. The peak is now located at
 10$^{15.7}$ eV and is higher by a factor of about 7. The contribution
 of IRB is now smaller than that of MBR ($\alpha$ = 2) only by about
 30\%. The highest curve in Fig.~\ref{ir_only} shows the IRB
 contribution for $\alpha$ = 2.5 and cosmological evolution with
 $n$ = 4 and then constant to $z$=10 followed by an exponential 
 decrease. The location of the peak does not change but its
 magnitude increases by almost a factor of three. It is now 50\%
 higher than the `standard' MBR generated cosmological neutrinos.
 It is obviously not correct to compare fluxes obtained with
 different assumptions for the cosmological evolution and we 
 do it only to have a feeling for the magnitude of the neutrino
 fluxes.
 The $\alpha$ = 2.5 spectra decrease the flux of cosmological
 neutrinos of energy above 10$^{19}$ eV.

 Both the spectral shape and the cosmological
 evolution of the UHECR sources affect the contribution of
 the IRB to the cosmological neutrino flux. The most important factor,
 however, is the shape of the injection spectrum. It is worth to note
 that the maximum proton energy at acceleration does not affect
 the IRB generated fluxes, since they are due mostly to protons of
 energy below 10$^{20}$ eV, as can be observed in Fig.~\ref{nuprod} 

  
 At energy about 3$\times$10$^{18}$ eV the cosmological fluxes
 of $\nu_\mu + \bar{\nu}_\mu$ are very close to the limit for
 source neutrinos. The reason is simple - in propagation from 
 large distances protons lose almost all of their energy in
 interactions on MBR. An interesting feature is the flux of 
 $\bar{\nu}_e$ (not shown), which peaks at energy about
 3$\times$10$^{15}$ eV. The origin of this flux is neutron decay,
 and a small  $\bar{\nu}_e$ flux is generated in neutron interactions
 on MBR.
 
  The cosmological evolution of the sources (n=3) increases the 
 fluxes by about a factor of five compared to a no-evolution
 scenario. The increase, however, is not energy independent~\cite{ESS01}.
 The highest energy neutrinos are generated at small redshifts.
 The low energy neutrinos come from high redshifts because of two
 reasons: the threshold energy of protons for photoproduction
 interaction decreases, and the generated neutrinos are further
 redshifted to the current epoch. The standard flux
 ($\alpha$=2.0, $n$=3) would generate about 0.4 neutrino induced
 showers  per km$^3$ year in the IceCube~\cite{IceCube} neutrino
 detector and 0.9 events with energy above 10$^{19}$ eV in the
 Auger\cite{Auger} observatory (for target mass of 30 km$^3$ of
 water) assuming that at  arrival at Earth the flavor ratio
 $\nu_e : \nu_\mu : \nu_\tau$  is 1:1:1 because of neutrino oscillations.
 It is difficult to estimate the rate in EUSO~\cite{EUSO} because of
 its yet unknown energy threshold.  These events come from the NC
 interactions of all neutrinos, CC interactions of $\nu_e$, the
 hadronic ($y$) part of the CC interactions of muon and tau neutrinos
 and from $\tau$ decay. Although very prominent,
 the Glashow resonance  does not produce high rate of events because
 of its narrow  width. Ice Cube should also detect very energetic
 muons with a comparable rate which is difficult to predict without
 detector Monte Carlo simulations.

\begin{figure}[thb]
\centerline{\includegraphics[width=3.truein]{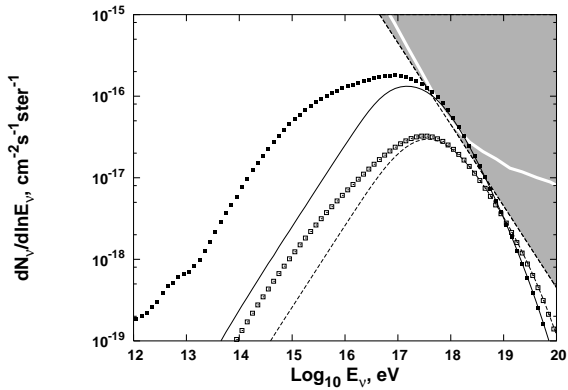}}
%
\caption{Comparison of the cosmological neutrino fluxes
with the Waxman\&Bahcall limit, which is given as an shaded area
for the `standard' power law injection spectrum and cosmological
evolution. The thick white line shows the limit
derived in Ref.~\protect\cite{MPR}. The dashed line shows the
flux of cosmological neutrinos generated in interactions on MBR 
for the `standard' parameters, and the solid one - for \protect$\alpha$
= 2.5 and $n$ = 4.  The squares show the fluxes generated on the
total photon background shown in Fig.~\ref{irbmbr}: the open squares
are for the `standard' parameters and the full ones - for 
\protect$\alpha$ = 2.5 and \protect$n$ = 4.    
\label{spectra_all}
}
\end{figure}

  Changing the proton injection spectrum to a power law with
 $\alpha$ = 2.5 moves the maximum of the cosmological neutrino
 flux to lower energy and increases the contribution of the
 interactions on IRB. At the same time the flux of higher
 energy cosmological neutrinos decreases. The shower event
 rates in IceCube and Auger become 0.44 and 0.31 respectively.

  Assuming a stronger source evolution, $(1+z)^4$ makes a big
 difference in the expected fluxes. With a power law source spectrum
 with $\alpha$ = 2.5 it generates 1.2 events in IceCube and
 0.66 events in Auger. The cosmological neutrino spectrum for 
 $\nu_\mu + \bar{\nu}_\mu$ is shown with full squares in 
 Fig.~\ref{spectra_all}. The contribution of interactions
 on MBR is shown with a solid line.


  The biggest uncertainty in these results, which is not listed
 above, is the cosmological evolution of the infrared/optical
 background. The estimates above assume that it is the same as
 of MBR, i.e. that the IRB was fully developed at $z$=8, which is
 the limit of the redshift integration. This does not seem to
 be a realistic assumption, although models of the IRB
 emission~\cite{HST03} predict very strong evolution of the far
 infrared emission, especially between redshifts of 10 to 100.

  The maximum proton energy at acceleration $E_{max}$ is unknown,
 but having in mind the highest energy Fly's Eye shower of
 3$\times$10$^{20}$ eV one should expect that astrophysical
 sources accelerate protons at least to 10$^{21}$ eV. 
 The injection spectrum is also not very well determined since
 the result of proton propagation depends on the UHECR source
 distribution. Attempts to derive the injection spectrum in the
 case of isotropic homogeneous source distribution end up with 
 injection spectra not flatter than $E^{-2.4}$ power
 law~\cite{BD04,Sta03}. . The extreme case is developed by
 Berezinsky et al.~\cite{BerGazGri02} who derive  an $\alpha$=2.7
 injection spectrum. 

 The luminosity required for the explanation of the observed events
 above 10$^{19}$ eV grows with the spectral index, and in the 
 case of Berezinsky et al. becomes 4.5$\times$10$^{47}$
 erg Mpc$^{-3}$yr$^{-1}$. Such steep spectrum would generate only
 a small event rate for neutrinos above 10$^{19}$ eV and would 
 enhance the IRB contribution.
 
  Expressed in terms of $(1 + z)^n$ the cosmological evolution of
 different astrophysical objects is observed to be between $n$ = 3
 and 4. A strong evolution with $n$ = 4, as used above, may be too
 optimistic, but not entirely out of range. As seen from
 Fig.~\ref{spectra_all}  strong cosmological evolution does not only
 increase the total flux, but moves the peak of the cosmological
 neutrino spectrum  to somewhat lower energy.

  Finally, the cosmic ray source luminosity, which was normalized to
 the flux of UHECR at 10$^{19}$ by Waxman~\cite{Wax95} could easily
 be higher or lower by half an order of magnitude. One can then assume
 a pessimistic IceCube shower event rate of 0.1 event per km$^3$yr 
 and an optimistic rate of 4-5 events. 

  It is obvious that a detailed calculation of the flux of cosmological
 neutrinos should include the interactions on the infrared background.
 We plan to do that with a better model of the IRB cosmological evolution
 and describe the calculation in more detail in a forthcoming paper.
 One should also keep in mind that if the UHECR sources are located in
 regions of high infrared and optical photon density, the fluxes of
 source neutrinos could increase. The effect may be much stronger
 if $~$10$^{19}$ eV and lower energy protons are contained in the
 region by high magnetic fields.
  
{\bf Acknowledgments} 
 The author acknowledges fruitful discussions with P.L.~Biermann,
 P.~Blasi, A.~Franceschini, D.~Seckel and S.~Yoshida. This research
 is supported in part by NASA Grant NAG5-10919.


\begin{thebibliography}{999}
\bibitem{IceCube} The current status of the ICECUBE project is displayed
 at pheno.physics.wisc.edu/icecube/.
\end{thebibliography}
\end{document}